\documentclass{article}
\usepackage[utf8]{inputenc}
\usepackage{amsmath,graphicx}

\begin{document}

\hspace*{12 cm}LA-UR-22-22687

\bigskip

\begin{center}
\bf \large LANSCE-PSR Short-Pulse Upgrade for Improved Dark Sector Particle Searches with the Coherent Captain Mills Experiment
\end{center}

\bigskip

\begin{center} E.-C. Huang, A.J. Hurd, W.C.~Louis, S.V. Milton, N.A.~Moody, P.~deNiverville,\\ C.E.~Taylor, R.T.~Thornton, R.G.~Van de Water$^{*}$

\it Los~Alamos~National~Laboratory,~Los~Alamos,~NM~87545,~USA
\end{center}


\begin{center}
S.G.~Biedron$^{*}$, M.~Fazio, S.I.~Sosa, T.J. Schaub

\it University~of~New~Mexico,~Albuquerque,~NM~87131,~USA
\end{center}


\begin{center}
J.W. Lewellen

\it SLAC National Accelerator Laboratory, Menlo Park, California 94025-7015
\end{center}


\bigskip

\begin{abstract}
  Proton beam dumps are prolific sources of charged and neutral pions, enabling a powerful technique to search for dark matter, axions, sterile neutrinos, tests of short baseline anomalies, and precision measurements of coherent nucleus scattering neutrinos (CEvNS). The Lujan neutron elastic scattering center at the Los Alamos Neutron Science Center (LANSCE) consists of an 800-MeV, short-pulse, 100-kW proton and spallation neutron source where such searches are ongoing with the Coherent CAPTAIN Mills (CCM) 10-ton, liquid argon detector. The employment of fast timing coincidence of the beam with the detector is used to identify signals and reject background. The current beam time width is 300\,ns with an intensity of $3.1 \times 10^{13}$ protons per pulse at 20\,Hz. With upgrades to the Proton Storage Ring (PSR), the beam time width may be compressed to 30\,ns with minimal intensity loss, allowing an increase in the signal to background (S/B) of more than 100 and an increase in the sensitivity for dark matter and sterile neutrino searches of an order of magnitude. This can be achieved with PSR accelerator upgrades on a time scale of a few years and at a modest cost.
 
\end{abstract}

\smallskip
$^{*}$\textit{Contact: biedron@unm.edu, vdwater@lanl.gov}

\section{Physics Motivation and Setup}

Our search is motivated by recent theoretical work that has shown that sub-GeV dark matter candidates can interact with ordinary matter through new light mediator particles \cite{DS, CVision, BRN}. Concurrent with this, the MiniBooNE experiment running at the  FNAL Booster Neutrino Beamline (BNB) carried out a special beam dump run that suppressed neutrino-produced backgrounds while enhancing the search for sub-GeV dark matter via neutral current scattering, resulting in significant new sub-GeV dark matter limits \cite{MB-DM}. This result clearly demonstrated the unique and powerful ability to search for dark matter with a beam dump neutrino experiment, especially with stopped pion sources \cite{SP}. Furthermore, at the lowest energies, CEvNS events were recently observed \cite{ch1, ch2}, and measuring the absolute CEvNS rate can provide a test of Non-Standard Interactions (NSI) \cite{NSI}. In addition, measuring the prompt muon neutrino CEvNS rate as a function of distance can test sterile neutrino interpretation models of the LSND and MiniBooNE anomalies \cite{PIPII-BD}. 
Recent analyses have demonstrated that tests for axions at the $\sim$MeV scale are ideally done at stopped pion sources \cite{axion}, and new models that connect dark sector particles to meson decays can explain the MiniBooNE short baseline anomaly and can be tested as well \cite{DSpion}.

 The Lujan tungsten spallation target is fed by 800-MeV protons from the Proton Storage Ring (PSR), providing a high-average power ($\sim$100 kW), 20\,Hz, short-pulse ($<$300\,nsec) source of moderated neutrons, neutrinos, and other potentially beam-related exotic particles, such as dark matter and axions. Neutrons are the main beam-related background for a detector placed 20\,m from the proton target. In addition, steady-state random backgrounds from cosmic rays, $^{39}$Ar decay, and external gamma-rays from long-lived activation are present - totaling $\sim$20\, kHz before analysis cuts. A shortened beam pulse can reduce these backgrounds significantly while not impacting the moderated neutron flux for material diffraction experiments elsewhere in the facility.
 
The Coherent CAPTAIN-Mills (CCM) detector is a 10-ton, liquid argon (LAr) fast detector instrumented with 120 photomultiplier tubes and 500\,MHz readout electronics. In 2019 it was built and ran for four months at Lujan center and has produced initial results on sub-GeV dark matter searches \cite{Aguilar-Arevalo:2021sbh, CCM:2021leg}. Measurements demonstrated that the detector can achieve $\sim$1 ns timing resolution, $\sim$20 cm spatial resolution, and an energy threshold of 50 keV (nuclear recoil). However, analysis has identified that LAr purification is required to remove scintillation light-absorbing O$_2$ and H$_2$O impurities to achieve 10-20 keV thresholds. The new and improved CCM200 detector (with 200 inner PMTs) was built with a purification system and began taking beam and calibration data in 2021. Initial results on detector performance are encouraging.

\begin{figure}[ht]
 \centerline{ 
 \includegraphics[width=0.70\textwidth]{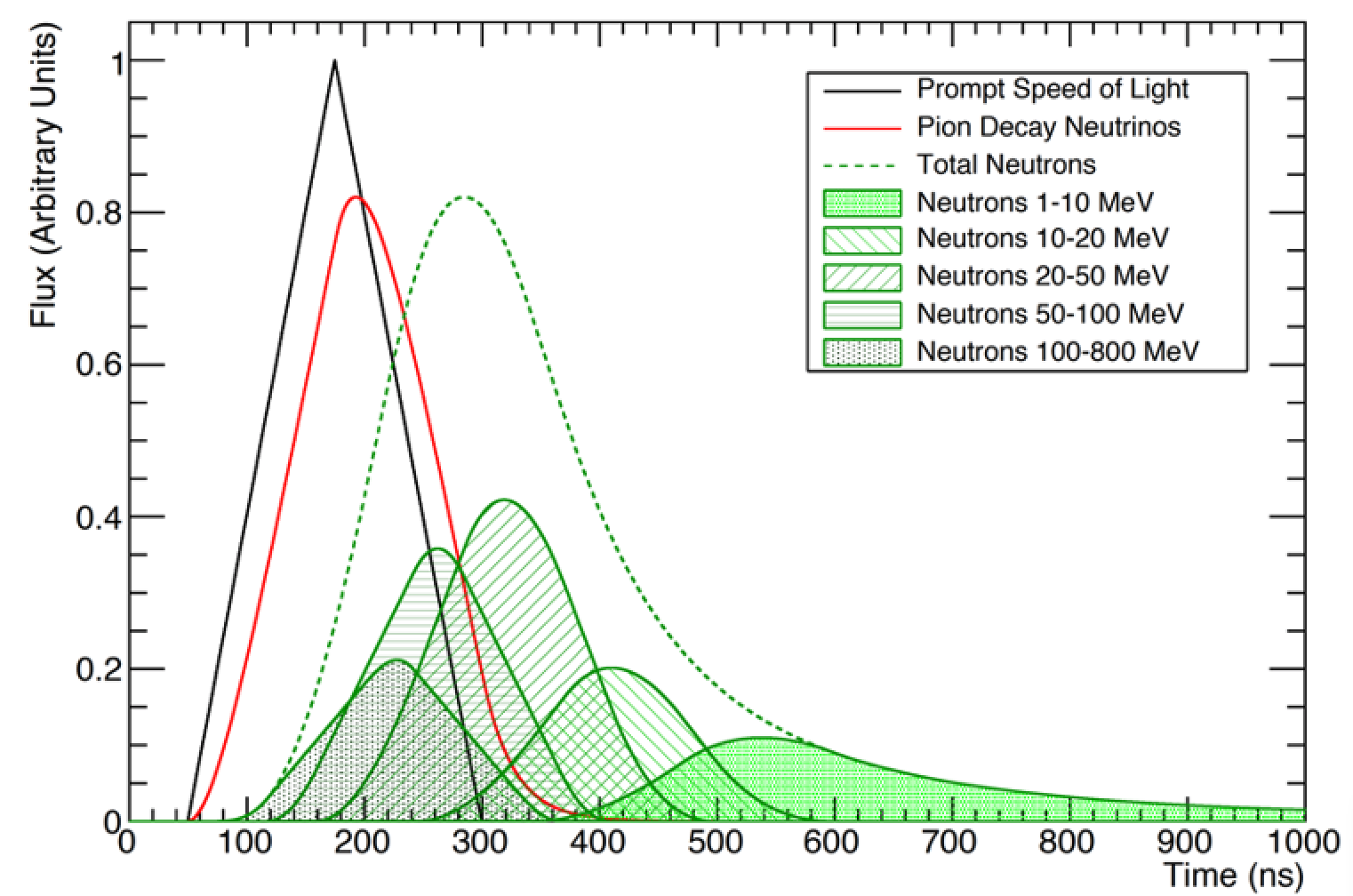}\hspace*{0.3cm}
 }
\caption{\footnotesize The expected timing distribution relative to the beam for various signals and backgrounds for the CCM detector located 20\,m from the Lujan target. Measurements of neutron time of flight at CCM using small portable detectors determined the fastest neutrons to be less than 20\,MeV, which implies a $\sim$200\,ns window where speed of light particles arrive free of beam neutrons. }
\label{fig1}
\end{figure}

\begin{figure}[ht]
 \centerline{ 
 \includegraphics[width=0.70\textwidth]{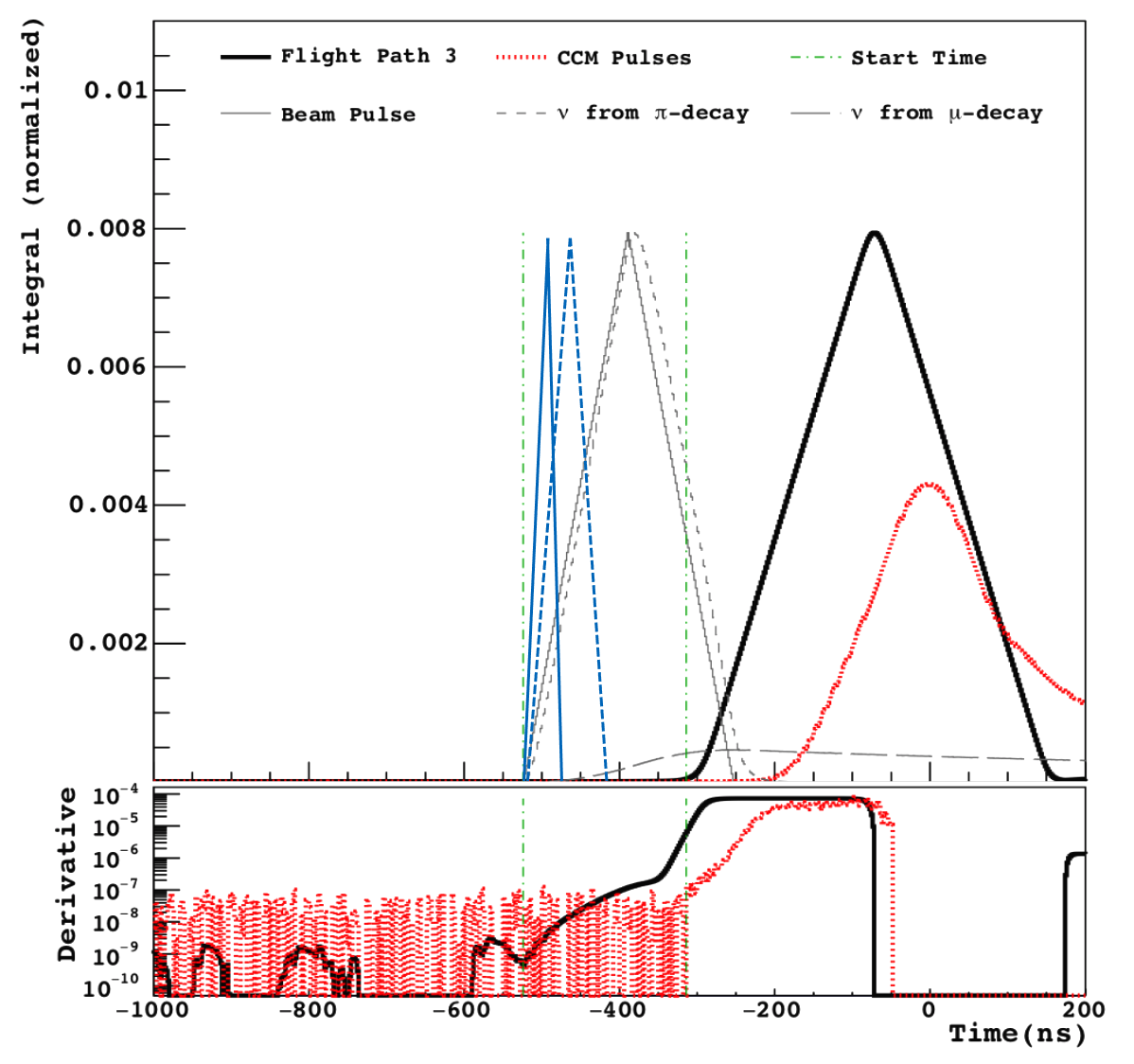}\hspace*{0.3cm}
 }
\caption{\footnotesize The measured absolute beam timing of the proton on target gamma-flash (FP3) relative to the CCM data from the Lujan source \cite{CCM:2021leg}. This demonstrates that there exists a $\sim$200\,ns window where neutrinos and other speed of light particles arrive free of subluminal neutrons. The light black triangle is the current 300\,ns beam pulse profile, while the proposed beam profiles are shown as dashed blue 100\,ns, and solid blue 30\,ns. The shorter beam profile will increase signal efficiency and significantly reduce steady-state random backgrounds. }
\label{fig2}
\end{figure}

Mono-energetic muon neutrinos come from charged pion decays at rest with a decay time of 26\,ns. This results in prompt muon neutrinos being mostly emitted in time with the beam. In addition, dark matter and axion particles that result from $\pi^0$ decay ($8.4 \times 10^{-8}$\,ns) are produced in time with the fast pulsed beam and can scatter off argon nuclei. These particles are more energetic than muon neutrinos from charged pion decay at rest and can be separated with a 50\,keV energy cut. The Lujan beam's unique and powerful timing features enable a relatively pure sample of muon neutrinos, dark matter, and axion candidates to be isolated in a region free of beam-related neutron backgrounds. Figure \ref{fig1} shows the anticipated timing features of the beam and various sources of signals and backgrounds. During normal operations, the PSR at LANSCE delivers a triangular proton pulse to the spallation target with a 300\,ns full base width. Detailed timing measurements \cite{CCM:2021leg} have determined that moderated neutrons arrive through the extensive bulk shielding at the CCM detector about $\sim$200\,ns after the arrival of speed of light particles, as shown in Figure \ref{fig2}. This window means that CCM can measure over 67\% of the prompt signal before the large wave of beam neutrons arrives. 
 
\begin{figure}[ht]
 \centerline{ 
 \includegraphics[width=0.65\textwidth]{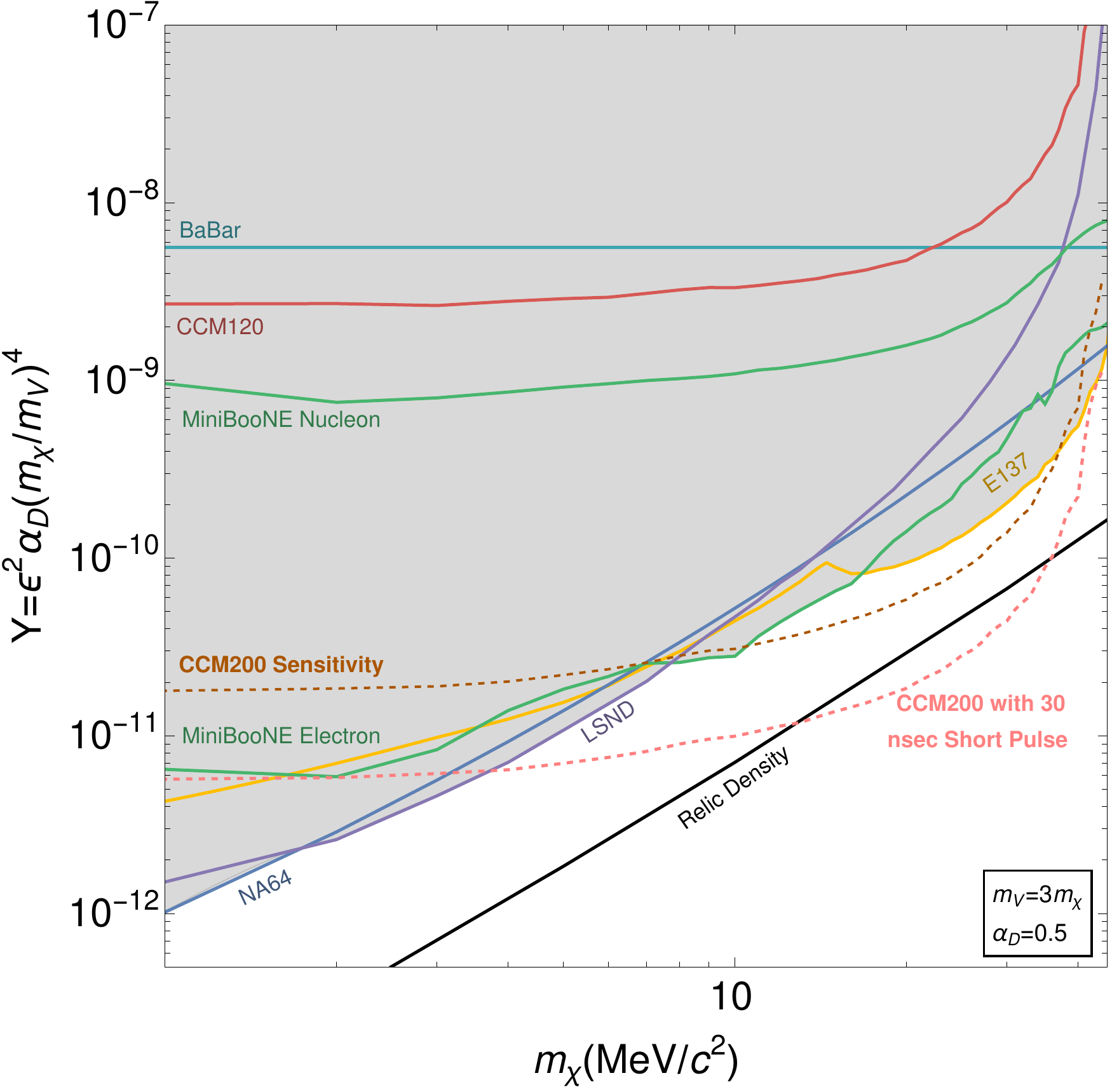}\hspace*{0.3cm}
 }
\caption{\footnotesize The dotted dark red line shows the expected CCM200 dark matter full fit sensitivity with a 10 keV threshold for a three-year run ($2.25\times10^{22}$~POT) based on estimates in \cite{CCM:2021leg}. The dotted pink line could be achieved with the 30\,ns Lujan short pulse upgrade to further reduce random backgrounds and begins to probe the relic density limits above 10 MeV.}
\label{fig3}
\end{figure}

To achieve the best efficiency, light dark sector particles and prompt neutrinos must arrive before the beam neutron background. If the proton beam pulse width can be shortened to less than 100\,ns, then about a 30\% improvement can be achieved in efficiency over the current situation. More importantly, the shorter beam pulse also reduces the steady-state random backgrounds by one-third (100/300), which delivers a combined four-fold signal-to-background (S/B) improvement for the same charge on target. If the proton beam could be reduced to 30\,ns while still maintaining the same charge per pulse, then a S/B of $\sim 1.3 / (30/300)$, or 13, can be achieved. In addition, the shorter signal window enables robust particle identification to suppress $^{39}$Ar backgrounds using the singlet to triplet light separation technique achieving up to another factor of ten or more separation. This results in a combined S/B of greater than 100. An example of the anticipated improvement with the short pulse upgrade for dark matter searches is shown in Figure \ref{fig3}.

While not directly related to the CCM experiment, some Lujan primary neutron spallation source users may also benefit from such a reduction in pulse length. For example, time-of-flight experimenters have a figure of merit that is proportional to the charge delivered on target and inversely proportional to the square of the bunch length. Reducing the bunch length, even at the expense of some charge loss, could significantly improve the performance of their experiments.

\section{LANSCE-PSR Short-Pulse Upgrades}

While shorter proton pulses in the PSR are desirable in neutron resonance experiments and would profoundly affect the CCM experiment, achieving sub-100\,ns pulse lengths with full beam intensity is hardly straightforward. Here we outline a few pathways for experimental tests to help define the requirements for PSR machine improvements.
In this effort, we are leveraging many theoretical and experimental studies that explored the facets of reaching better timing, higher currents, higher peak currents, and shorter pulses in the PSR \cite{ac1, ac2}. Instabilities play into the limitations when operating in short-pulse mode. For instance, the PSR can suffer from an electron cloud (e-p) instability \cite{ac3} and initial experiments have been performed for active damping of the instability through the implementation of an analog vertical feedback system \cite{ac4, ac5} and the implementation of heated ferrites. However, the ferrites actually induce a longitudinal microwave instability when attempting to reach short pulse mode. There is, therefore, a balance between the mitigation of the transverse and longitudinal instabilities, particularly in the quest to achieve short pulses.
	
We believe an active transverse feedback system in both the vertical and horizontal plane to control the (e-p) instability must be considered to achieve significantly shorter bunch lengths. This would be supplemented by the use of novel magnetic materials that would replace the existing ferrites. These new materials will have properties that do not inadvertently induce a longitudinal (microwave) instability \cite{ac6} \cite{IPAC2021_TAYLOR}. A longitudinal feedback system might be necessary should the threshold of the microwave instability occur at the higher peak currents we require.
In addition to the above, using a second harmonic RF cavity might also prove advantageous to achieve the shortest bunch lengths. Such a cavity could be tuned to cancel the fourth-order term in the $\cos(\theta)$ expansion of the RF field, creating a broader region in longitudinal phase space where the synchrotron period as a function of amplitude is held constant. This would mitigate nonlinear rotation that creates longitudinal tails.

\section{Conclusion}

It is expected that once designed, the PSR upgrades to achieve 30\,ns timing could be implemented in 2-3 years and for under \$5M.   The risks are anticipated to be medium and, if successful, would have a significant impact ($\sim \mathcal{O}(10)$ sensitivity) on the physics capability of the CCM experiment to search for dark matter, axions, sterile neutrinos, and CEvNS.

In general, when the upper limits of beam intensity have been reached, shifting efforts to increase the instantaneous accelerator power by reducing the beam pulse width can yield positive results. This can lead to a dramatic, cost-effective increase in sensitivity to neutrino and dark sector physics.

\end{document}